\documentclass[reprint, 10pt, amsmath, amssymb, floatfix, aps, prl,  superscriptaddress, nofootinbib, nobibnotes,twocolumn]{revtex4-1}

\pdfoutput=1
\usepackage{graphicx}
\graphicspath{{figures/}}
\usepackage[utf8]{inputenc}
\usepackage{amsmath}
\usepackage{amsfonts}
\usepackage{amssymb}
\usepackage{multirow}
\usepackage{CJKutf8}
\usepackage{color}
\usepackage[colorlinks=true, linkcolor=red, citecolor=blue]{hyperref}
\usepackage[capitalise]{cleveref}
\usepackage{acro}
\usepackage{adjustbox}
\usepackage{array}
\usepackage{natbib}
\usepackage{dblfnote}
\DFNalwaysdouble
\usepackage{slashed}
\usepackage{dcolumn}
\usepackage{hyperref}
\usepackage{geometry}
\usepackage{subfig} 
\usepackage{overpic}
\usepackage{inputenc}
\usepackage{caption}
\def\({\left(}
\def\){\right)}
\def\[{\left[}
\def\]{\right]}
\def\be{\begin{eqnarray}}
\def\ee{\end{eqnarray}}

\DeclareAcronym{GW}{
  short = GW ,
  long = gravitational wave ,
  short-plural = s 
}
\DeclareAcronym{LIGO}{
  short = LIGO ,
  long = Laser Interferometer Gravitational-wave Observatory ,
  short-plural = 
}
\DeclareAcronym{LISA}{
  short = LISA ,
  long = Laser Interferometer Space Antenna ,
  short-plural =  
}
\DeclareAcronym{SKA}{
  short = SKA ,
  long = Square Kilometre Array ,
  short-plural =  
}  
  
\DeclareAcronym{SNR}{
	short = SNR ,
	long = signal-to-noise ratio ,
	short-plural = 
}

\DeclareAcronym{PTA}{
	short = PTA ,
	long = pulsar timing array ,
	short-plural = 
}

\DeclareAcronym{FLRW}{
  short = FLRW ,
  long = Friedmann-Lemaitre-Robertson-Walker ,
  short-plural =  
}

\DeclareAcronym{SIGW}{
	short = SIGW ,
	long = scalar induced gravitational wave ,
	short-plural =  s
}

\DeclareAcronym{PBH}{
	short = PBH ,
	long = primordial black hole ,
	short-plural =  s
}

\DeclareAcronym{CMB}{
	short = CMB ,
	long = cosmic microwave background ,
	short-plural =  
}
\DeclareAcronym{DM}{
	short = DM ,
	long = dark matter ,
	short-plural =  
}

\DeclareAcronym{SGWB}{
	short = SGWB ,
	long = stochastic gravitational	wave background ,
	short-plural =  s
}

\DeclareAcronym{LSS}{
	short = LSS ,
	long = large scale structure ,
	short-plural =  
}

\DeclareAcronym{RD}{
	short = RD ,
	long = radiation-dominated ,
	short-plural =  
}

\begin{document}

\title{New constraints on primordial non-Gaussianity from missing two-loop contributions of scalar induced gravitational waves
}

\author{Zhe Chang} 
\affiliation{Institute of High Energy Physics, Chinese Academy of Sciences, Beijing 100049, China}
\affiliation{University of Chinese Academy of Sciences, Beijing 100049, China}

\author{Yu-Ting Kuang} 
\affiliation{Institute of High Energy Physics, Chinese Academy of Sciences, Beijing 100049, China}
\affiliation{University of Chinese Academy of Sciences, Beijing 100049, China}

\author{Di Wu} 
\affiliation{Institute of High Energy Physics, Chinese Academy of Sciences, Beijing 100049, China}
\affiliation{University of Chinese Academy of Sciences, Beijing 100049, China}

\author{Jing-Zhi Zhou} 
\email{zhoujingzhi@ihep.ac.cn}
\affiliation{Institute of High Energy Physics, Chinese Academy of Sciences, Beijing 100049, China}
\affiliation{University of Chinese Academy of Sciences, Beijing 100049, China}

\author{Qing-Hua Zhu} 
\affiliation{Department of Physics, Chongqing University, Chongqing 401331, China}

\begin{abstract}
We analyze the energy density spectrum of \acp{SIGW} using the NANOGrav 15-year data set, thereby constraining the primordial non-Gaussian parameter $f_{\mathrm{NL}}$. For the first time, we calculate the seventeen missing two-loop diagrams proportional to $f_{\mathrm{NL}}A_{\zeta}^3$ that correspond to the two-point correlation function $\langle h^{\lambda,(3)}_{\mathbf{k}} h^{\lambda',(2)}_{\mathbf{k}'} \rangle$ for local-type primordial non-Gaussianity. The total energy density spectrum of \acp{SIGW} can be significantly suppressed by these two-loop diagrams. If \acp{SIGW} dominate the \acp{SGWB} observed in \ac{PTA} experiments, the parameter interval $f_{\mathrm{NL}}\in [-5,-1]$ is notably excluded based on NANOGrav 15-year data set. After taking into account abundance of \acp{PBH} and the convergence of the cosmological perturbation expansion, we find that the only possible parameter range for $f_{\mathrm{NL}}$ might be $-1\le f_{\mathrm{NL}}< 0$.
\end{abstract}

\maketitle
\acresetall

\emph{\textbf{Introduction.}}---Recently, the \ac{PTA} collaborations NANOGrav \cite{NANOGrav:2023gor,NANOGrav:2023hvm}, EPTA \cite{EPTA:2023fyk}, Parkers PTA \cite{Reardon:2023gzh}, and the China PTA \cite{Xu:2023wog} have reported positive evidence for an isotropic, stochastic background
of \acp{GW} within the nHz frequency range. Numerous potential sources contribute to the \acp{SGWB}. For standard astrophysical sources, the \ac{PTA} signal is predominantly attributed to supermassive black hole binaries \cite{Middleton:2020asl,NANOGrav:2020spf}. In addition, the data might also have a cosmological origin, such as first-order phase transitions \cite{Fujikura:2023lkn,Addazi:2023jvg,Jiang:2023qbm,Xiao:2023dbb,Wu:2023hsa,He:2023ado}, cosmic strings \cite{Ellis:2020ena,Ellis:2023tsl,Lazarides:2023ksx,Yamada:2023thl,Qiu:2023wbs}, and \acp{SIGW} \cite{Vaskonen:2020lbd,DeLuca:2020agl,Balaji:2023ehk,Franciolini:2023pbf,Balaji:2023ehk,You:2023rmn,Zhao:2023joc,Wang:2023ost,Zhu:2023faa,Yu:2023jrs,Chang:2023ist}. Both astrophysical and cosmological sources could play a crucial role in shaping the \acp{SGWB}. This Letter considers the possibility that recent \acp{PTA} data can be explained by \acp{GW} induced by non-Gaussian primordial curvature perturbations.

In recent years, research on \acp{SIGW} has received widespread attention \cite{Ananda:2006af,Baumann:2007zm,Domenech:2021ztg}. The \acp{SIGW} are produced by a higher order effect that emerges from scalar perturbations re-entering the horizon after inflation. Since the constraints of primordial curvature
perturbations on small scales ($\lesssim$1 Mpc) are significantly weaker than those on
large scales \cite{Planck:2018vyg,Abdalla:2022yfr,Bringmann:2011ut}, the \acp{SIGW} can be used as a probe of the small-scale primordial power spectrum to help us understand the specific properties of quantum fluctuations on small scales during the inflationary period \cite{Saito:2008jc,Inomata:2018epa}.

In cosmological perturbation theory, the cosmology perturbations can be decomposed as scalar, vector, and tensor perturbations \cite{Malik:2008im,Mukhanov:1990me,Kodama:1984ziu}. The tensor perturbations, known as \acp{GW} on \ac{FLRW} spacetime, can be written as $h^{\lambda}_{ij}=h^{\lambda,(1)}_{ij}+\frac{1}{2}h^{\lambda,(2)}_{ij}+\frac{1}{6}h^{\lambda,(3)}_{ij}+\cdot\cdot\cdot$, where $h^{\lambda,(n)}_{ij}$ is the $n$-th order gravitational wave. For $n=1$, $h^{\lambda,(1)}_{ij}$ is known as the primordial GWs; for $n>1$, $h^{\lambda,(n)}_{ij}$ are higher order GWs induced by lower order perturbations. The energy density spectra of the GWs can be calculated in terms of the two-point function of GWs, $\langle h^{\lambda}_{ij}(\mathbf{k}) h^{\lambda'}_{ij}(\mathbf{k}') \rangle$. Substituting the cosmological perturbation expansion of GWs into the two-point function, we obtain $\langle h^{\lambda}_{ij}(\mathbf{k}) h^{\lambda'}_{ij}(\mathbf{k}') \rangle=\sum_{n,m=1}^{\infty}\frac{1}{n!m!} \langle h^{\lambda,(n)}_{ij} h^{\lambda',(m)}_{ij} \rangle$. If we neglect the effect of primordial tensor perturbation, namely $h^{\lambda,(1)}_{ij}\approx 0$ on small scale, then $ \langle h^{\lambda,(1)}_{ij} h^{\lambda',(n)}_{ij} \rangle=0$ for arbitrary $h^{\lambda,(n)}_{ij}$. In this case, the lowest order contribution of the two-point function of GWs is $\langle h^{\lambda,(2)}_{ij} h^{\lambda',(2)}_{ij} \rangle$, where $h^{\lambda,(2)}_{ij}$ is known as the second order \ac{SIGW}.
The semianalytic calculation of the second order SIGWs was presented in Ref.~\cite{Kohri:2018awv}, and has been widely used in research related to SIGWs and primordial black holes (PBHs) \cite{Wang:2019kaf,Byrnes:2018txb,Inomata:2020lmk,Ballesteros:2020qam,Lin:2020goi,Chen:2019xse,Cai:2019elf,Cai:2019jah,Ando:2018qdb,Di:2017ndc,Gao:2021vxb,Changa:2022trj,Zhou:2020kkf,Cai:2021wzd,Chen:2022dah,Chang:2022vlv}. Furthermore, the studies on SIGWs have also extended to gauge issue \cite{Hwang:2017oxa,Yuan:2019fwv,Inomata:2019yww,DeLuca:2019ufz,Domenech:2020xin,Chang:2020tji,Ali:2020sfw,Lu:2020diy,Tomikawa:2019tvi,Gurian:2021rfv,Uggla:2018fiy,Ali:2023moi}, primordial non-Gaussianity \cite{Cai:2018dig,Atal:2021jyo,Zhang:2020uek,Yuan:2020iwf,Davies:2021loj,Rezazadeh:2021clf,Kristiano:2021urj,Bartolo:2018qqn,Adshead:2021hnm,Li:2023qua,Li:2023xtl}, damping effect \cite{Mangilli:2008bw,Saga:2014jca,Zhang:2022dgx,Yuan:2023ofl}, different epochs of the Universe \cite{Papanikolaou:2020qtd,Domenech:2020kqm,Domenech:2019quo,Inomata:2019zqy,Inomata:2019ivs,Assadullahi:2009nf,Witkowski:2021raz,Dalianis:2020gup,Hajkarim:2019nbx,Bernal:2019lpc,Das:2021wad,Haque:2021dha,Domenech:2020ssp,Domenech:2021and,Liu:2023pau}, modified gravity \cite{Papanikolaou:2021uhe,Papanikolaou:2022hkg,Tzerefos:2023mpe}, and third order SIGWs $\langle h^{\lambda,(3)}_{ij} h^{\lambda',(3)}_{ij} \rangle$ \cite{Zhou:2021vcw} in the past few years.

In this Letter, we neglect the effects of primordial GWs $h^{\lambda,(1)}_{ij}$. Then, the three lowest order contributions are provided by the following three two-point functions: $\langle h^{\lambda,(2)}_{ij} h^{\lambda',(2)}_{ij} \rangle$, $\langle h^{\lambda,(3)}_{ij} h^{\lambda',(2)}_{ij} \rangle$, and $\langle h^{\lambda,(3)}_{ij} h^{\lambda',(3)}_{ij} \rangle$. Two-point functions $\langle h^{\lambda,(2)}_{ij} h^{\lambda',(2)}_{ij} \rangle$ and $\langle h^{\lambda,(3)}_{ij} h^{\lambda',(3)}_{ij} \rangle$ represent the second and the third order SIGWs, respectively. The two-point function $\langle h^{\lambda,(3)}_{ij} h^{\lambda',(2)}_{ij} \rangle$ can be calculated in terms of a given five-point correlation function of primordial curvature perturbations $\langle\zeta_{\mathbf{k}-\mathbf{p}}\zeta_{\mathbf{p}-\mathbf{q}}\zeta_{\mathbf{q}}\zeta_{\mathbf{k}'-\mathbf{p}'}\zeta_{\mathbf{p}'}\rangle$. Obviously, the contributions of $\langle h^{\lambda,(3)}_{ij} h^{\lambda',(2)}_{ij} \rangle$ need to be considered for the local-type non-Gaussian primordial curvature perturbations $\zeta_{\mathbf{k}}=\zeta^g_{\mathbf{k}}+\frac{3}{5}f_{\mathrm{NL}}\int\frac{d^3n}{(2\pi)^{3/2}} \zeta^g_{\mathbf{k}-\mathbf{n}}\zeta^g_{\mathbf{n}} $.

In previous studies, the power spectra of GWs induced by non-Gaussian scalar perturbations have been found to originate from two components up to $(f_{\mathrm{NL}})^2$ order: the Gaussian part, $\langle h^{\lambda,(2)}_{ij} h^{\lambda',(2)}_{ij} \rangle$, which is proportional to $A_{\zeta}^2$, and the non-Gaussian part, $\langle h^{\lambda,(2)}_{ij} h^{\lambda',(2)}_{ij} \rangle$, which is proportional to $(f_{\mathrm{NL}})^2A_{\zeta}^3$ \cite{Adshead:2021hnm}. The non-Gaussian part corresponds to three kinds of two-loop diagrams in cosmological perturbation theory \cite{Adshead:2021hnm,Li:2023qua,Li:2023xtl}. In this Letter, we study the new contributions from the two-point function $\langle h^{\lambda,(3)}_{ij} h^{\lambda',(2)}_{ij} \rangle$ induced by local-type non-Gaussian scalar perturbation. The contributions of $\langle h^{\lambda,(3)}_{ij} h^{\lambda',(2)}_{ij} \rangle$ are proportional to $f_{\mathrm{NL}}A_{\zeta}^3$ at lowest order of $f_{\mathrm{NL}}$. In this case, the contributions of positive and negative $f_{\mathrm{NL}}$ values to the total energy density spectra of \acp{SIGW} are not degenerate. Furthermore, since the contributions of the new two-loop diagrams are negative when $f_{\mathrm{NL}}<0$, the total energy density spectrum of \acp{SIGW} can be greatly suppressed.

\emph{\textbf{Second and third order \acp{SIGW}.}}---The perturbed metric in the \ac{FLRW} spacetime with
Newtonian gauge takes the form
\begin{equation}
	\begin{aligned}
		&\mathrm{d}s^{2}=a^{2}\left(-\left(1+2 \phi^{(1)}+ \phi^{(2)}\right) \mathrm{d} \eta^{2}+ V_i^{(2)} \mathrm{d} \eta \mathrm{d} x^{i}+\right. \\
		&\left.\left(\left(1-2 \psi^{(1)}- \psi^{(2)}\right) \delta_{i j}+\frac{1}{2} h_{i j}^{(2)}+\frac{1}{6} h_{i j}^{(3)}\right)\mathrm{d} x^{i} \mathrm{d} x^{j}\right) \ ,
	\end{aligned}
\end{equation}
where $\phi^{(n)}$ and $\psi^{(n)}$$\left( n=1,2 \right)$ are first order and second order scalar perturbations. $h^{(n)}_{ij}$$\left( n=2,3 \right)$ are second order and third order tensor perturbations. $V_i^{(2)} $ is second order vector perturbation. The first order scalar perturbations in momentum space can be written as $\psi(\eta,\mathbf{k}) = \phi(\eta,\mathbf{k}) =\Phi_{\mathbf{k}} T_\phi(k \eta)= \frac{2}{3}\zeta_{\mathbf{k}} T_\phi(k \eta)$ where $\zeta_{\mathbf{k}}$ is the primordial curvature perturbation. The transfer function $ T_\phi(k \eta)$ in the \ac{RD} era is defined as $T_{\phi}(x)=\frac{9}{x^{2}}\left(\frac{\sqrt{3}}{x} \sin \left(\frac{x}{\sqrt{3}}\right)-\cos \left(\frac{x}{\sqrt{3}}\right)\right)$ \cite{Inomata:2020cck}. In momentum space, the equations of motion of higher order induced GWs during the \acp{RD} era are given by \cite{Kohri:2018awv}
\begin{equation}
	\begin{aligned}
		h_{i j}^{\lambda,(n)''}(\mathbf{k},\eta )+&\frac{2}{\eta} h_{i j}^{\lambda,(n)'}(\mathbf{k},\eta )+k^2 h_{i j}^{\lambda,(n)}(\mathbf{k},\eta )\\
		&=-12 \Lambda_{i j}^{l m}(\mathbf{k}) \mathcal{S}_{l m}^{(n)}(\mathbf{k},\eta ) \ , 
	\end{aligned}\label{eq:M}
\end{equation}
where $\Lambda_{i j}^{l m}=\mathcal{T}_i^l \mathcal{T}_j^m-\frac{1}{2} \mathcal{T}_{i j} \mathcal{T}^{l m}$ is the transverse and traceless operator in momentum space \cite{Zhou:2021vcw}, and $\mathcal{T}_i^l$ is defined as $\mathcal{T}_i^l=$ $\delta_i^l-k^lk_i/k^2$. $h_{i j}^{\lambda,(n)}(\mathbf{k},\eta )$$(n=2,3,\cdot\cdot\cdot)$ are $n$-th order SIGWs in momentum space. Eq.~(\ref{eq:M}) can be solved by the Green's function method, namely \cite{Kohri:2018awv}
\begin{equation}
	\begin{aligned}
		h^{\lambda,(n)}_{\mathbf{k}}(\eta)=\frac{12}{k \eta} \int^{\eta}_0 \mathrm{d} \tilde{\eta} \sin (k \eta-k \tilde{\eta}) \tilde{\eta} \mathcal{S}^{\lambda,(n)}_{\mathbf{k}}(\tilde{\eta}) \ ,
	\end{aligned}\label{eq:G}
\end{equation}
where we have defined $h_{\mathbf{k}}^{\lambda,(n)}(\eta)=\varepsilon^{\lambda, i j}(\mathbf{k}) h_{i j}^{(n)}(\mathbf{k}, \eta)$ and $\mathcal{S}_{\mathbf{k}}^\lambda(\eta)=-\varepsilon^{\lambda, l m}(\mathbf{k}) S_{l m}^{(3)}(\mathbf{k}, \eta)$. $\varepsilon_{i j}^\lambda(\mathbf{k})$ is the polarization tensor. 

For the second order \acp{SIGW}, we use the symbol $h^{(2)}_{\mathbf{k},\phi\phi}$ to represent the second order SIGWs sourced by two first order scalar perturbations \cite{Ellis:2016jkw}. The corresponding source term in momentum space is given by \cite{Kohri:2018awv}
\begin{equation}
	\begin{aligned}
		S_{\mathbf{k},\phi\phi}^{\lambda,(2)}(\eta)= \frac{4}{9}\int \frac{d^3 p}{(2 \pi)^{3 / 2}}  \varepsilon^{\lambda, l m}(\mathbf{k})p_l p_m \zeta_{\mathbf{k}-\mathbf{p}}  \zeta_{\mathbf{p}}  f_{\phi\phi}^{(2)}(u,v,x) \ ,
	\end{aligned}\label{eq:2s}
\end{equation}
where the coefficient $\frac{4}{9}$ comes from the definition of transfer function $ \phi(\eta,\mathbf{k})= \frac{2}{3}\zeta_{\mathbf{k}} T_\phi(k \eta)$. $f_{\phi\phi}^{(2)}(u,v,x)$ is given by
\begin{equation}
	\begin{aligned}
		f^{(2)}_{\phi\phi}(u,v,x)&=2 T_{\phi}(ux) T_{\phi}(vx) +\left(ux \frac{d}{d(ux)}T_{\phi}(ux)\right.\\
		&\left.+T_{\phi}(ux)\right)\left(vx \frac{d}{d(vx)}T_{\phi}(vx)+T_{\phi}(vx)\right)\ .
	\end{aligned}\label{eq:2f}
\end{equation}
Here, we have defined $|\mathbf{k}-\mathbf{p}|=u|\mathbf{k}|$, $|\mathbf{p}|=v|\mathbf{k}|$, and $x=|\mathbf{k}|\eta$. Substituting Eq.~(\ref{eq:2s}) and Eq.~(\ref{eq:2f}) into Eq.~(\ref{eq:G}), we can rewrite Eq.~(\ref{eq:G}) as
\begin{equation}
	\begin{aligned}
		h^{\lambda,(2)}_{\mathbf{k},\phi\phi}(\eta)= &\frac{4}{9}\int \frac{d^3 p}{(2 \pi)^{3 / 2}}  \varepsilon^{\lambda, l m}(\mathbf{k})p_l p_m \\
  & \times \zeta_{\mathbf{k}-\mathbf{p}}  \zeta_{\mathbf{p}} I_{\phi\phi}^{(2)}(u,v,x) \ ,
	\end{aligned}\label{eq:2h}
\end{equation}
where 
\begin{equation}
	\begin{aligned}
		I_{\phi\phi}^{(2)}(u,v,x)=\frac{4}{k^2} \int_0^x \mathrm{d} \bar{x}\left(\frac{\bar{x}}{x} \sin (x-\bar{x}) f_h^{(2)}(u,v,x)\right) 
	\end{aligned}\label{eq:2I}
\end{equation}
is known as the kernel function of second order SIGWs.

For the third order \acp{SIGW}, there are three kinds of source terms: $\mathcal{S}_1 \sim \phi^{(1)}\phi^{(1)}\phi^{(1)}$, $\mathcal{S}_2 \sim \phi^{(1)}h^{\lambda,(2)}_{\phi\phi}$, $\mathcal{S}_3 \sim \phi^{(1)}V^{\lambda,(2)}_{\phi\phi}$, and $\mathcal{S}_4 \sim \phi^{(1)}\psi^{(2)}_{\phi\phi}$ \cite{Zhou:2021vcw}. Then, the third order \acp{SIGW} can be written as \cite{Chang:2022nzu}
\begin{eqnarray}
	h^{\lambda,(3)}_{\mathbf{k}}(\eta)& &=h^{\lambda,(3)}_{\mathbf{k},\phi\phi\phi}(\eta)+h^{\lambda,(3)}_{\mathbf{k},\phi h_{\phi\phi}}(\eta)+h^{\lambda,(3)}_{\mathbf{k},\phi V_{\phi\phi}}(\eta) \nonumber\\
	& &+h^{\lambda,(3)}_{\mathbf{k},\phi \psi_{\phi\phi}}(\eta) \  ,
	\label{eq:H0}
\end{eqnarray}
where
\begin{align}
	h^{\lambda,(3)}_{\mathbf{k},\phi\phi\phi}&(\eta)=\int\frac{d^3p}{(2\pi)^{3/2}}\int\frac{d^3q}{(2\pi)^{3/2}}\varepsilon^{\lambda,lm}(\mathbf{k})q_m \nonumber\\
	&\times (p_l-q_l)\frac{8}{27} I_{\phi\phi\phi}^{(3)}(u,\bar{u},\bar{v},x)\zeta_{\mathbf{k}-\mathbf{p}} \zeta_{\mathbf{p}-\mathbf{q}} \zeta_{\mathbf{q}} \  ,
	\label{eq:H1}\\
	h^{\lambda,(3)}_{\mathbf{k},\phi h_{\phi\phi}}&(\eta)=\int\frac{d^3p}{(2\pi)^{3/2}}\int\frac{d^3q}{(2\pi)^{3/2}} \varepsilon^{\lambda, lm}(\mathbf{k})\Lambda_{lm}^{ rs}(\mathbf{p})\nonumber\\
	&\times
	q_rq_s\frac{8}{27}I^{(3)}_{\phi h_{\phi\phi}}(u,\bar{u},\bar{v},x)\zeta_{\mathbf{k}-\mathbf{p}} \zeta_{\mathbf{p}-\mathbf{q}} \zeta_{\mathbf{q}} \ ,
	\label{eq:H2} \\
	h^{\lambda,(3)}_{\mathbf{k},\phi V_{\phi\phi}}&(\eta)=\int\frac{d^3p}{(2\pi)^{3/2}}\int\frac{d^3q}{(2\pi)^{3/2}}\varepsilon^{\lambda,lm}(\mathbf{k})\mathcal{T}^r_{(m}(\textbf{p}) p_{l)}  \nonumber\\
	&\times  \frac{16p^s}{27p^2}q_rq_sI^{(3)}_{\phi V_{\phi\phi}}(u,\bar{u},\bar{v},x)\zeta_{\mathbf{k}-\mathbf{p}}\zeta_{\mathbf{p}-\mathbf{q}} \zeta_{\mathbf{q}} \ ,
	\label{eq:H3}\\
	h^{\lambda,(3)}_{\mathbf{k},\phi \psi_{\phi\phi}}&(\eta)=\int\frac{d^3p}{(2\pi)^{3/2}}\int\frac{d^3q}{(2\pi)^{3/2}}\varepsilon^{\lambda,lm}(\mathbf{k})p_lp_m \nonumber\\
	&\times
	\frac{8}{27}I^{(3)}_{\phi \psi_{\phi\phi}}(u,\bar{u},\bar{v},x)\zeta_{\mathbf{k}-\mathbf{p}} \zeta_{\mathbf{p}-\mathbf{q}} \zeta_{\mathbf{q}} \ .
	\label{eq:H4}
\end{align}
Here we have defined $|\mathbf{k}-\mathbf{p}|=uk$, $|\mathbf{k}-\mathbf{q}|=wk$,  $|\mathbf{p}-\mathbf{q}|=\bar{u}p=\bar{u}vk$, and $q=\bar{v}p=\bar{v}vk$. The explicit expressions of the third order kernel functions $I^{(3)}$ can be found in Ref.~\cite{Zhou:2021vcw}. 

\begin{figure*}[htbp]
    \centering
    \captionsetup{justification=centering,singlelinecheck=false}
\hspace{0.1\columnwidth}   
        \subfloat[]{\includegraphics[width=.4\columnwidth]{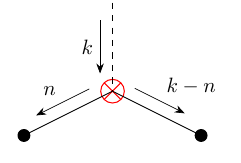}} 
        \hspace{0.1\columnwidth}
        \subfloat[]{\includegraphics[width=.7\columnwidth]{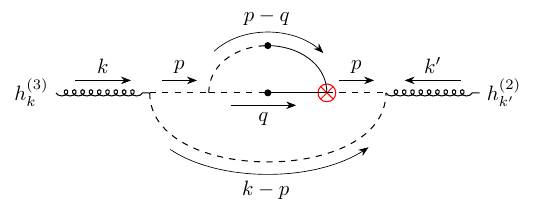}}
	\subfloat[]{\includegraphics[width=.7\columnwidth]{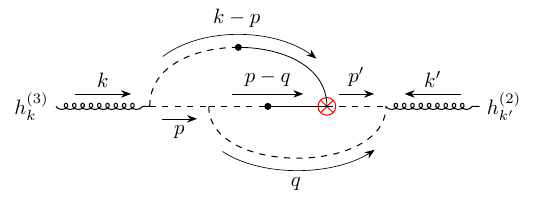}}  \\
	\subfloat[]{\includegraphics[width=.7\columnwidth]{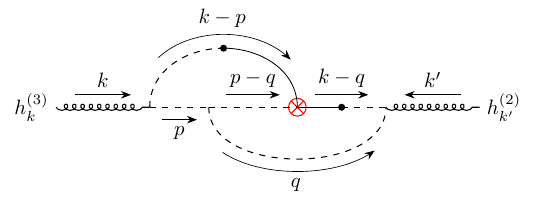}}
        \subfloat[]{\includegraphics[width=.65\columnwidth]{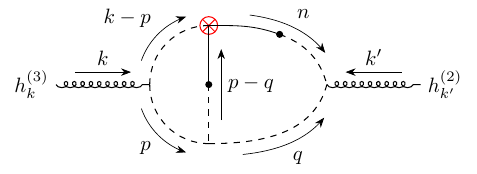}}
        \subfloat[]{\includegraphics[width=.7\columnwidth]{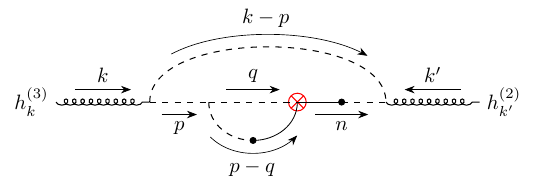}} 
        
\caption{\label{fig:FeynDiagLarge} The two loop contributions from Eq.~(\ref{eq:P}). The dashed and spring-like lines in the figure represent scalar and tensor perturbations, respectively. Figure (a) depicts the non-Gaussian vertex. }
\end{figure*}

\emph{\textbf{Missing two-loop corrections.}}---We consider the local type non-Gaussianity which can be expressed as a local perturbative expansion around the Gaussian primordial curvature perturbation. In momentum space, the primordial curvature perturbation can be rewritten as \cite{Domenech:2021ztg}
\begin{equation}
	\begin{aligned}
		\zeta^{\mathrm{ng}}_{\mathbf{k}}=\zeta_{\mathbf{k}}+\frac{3}{5}f_{\mathrm{NL}}\int\frac{d^3n}{(2\pi)^{3/2}} \zeta_{\mathbf{k}-\mathbf{n}}\zeta_{\mathbf{n}}  \ ,
	\end{aligned}\label{eq:ngk}
\end{equation}
where $\mathbf{n}$ is the three dimensional momentum variable. The two-point function $\langle h^{\lambda,(3)}_{\mathbf{k}}(\eta) h^{\lambda,(2)}_{\mathbf{k}'}(\eta) \rangle$  can be calculated in terms of Eq.~(\ref{eq:H0})--Eq.~(\ref{eq:H4}) and Eq.~(\ref{eq:2h}). For example, the two-point correlation function of $ h^{\lambda,(3)}_{\mathbf{k},\phi \psi_{\phi\phi}}(\eta)$ and $h^{\lambda',(2)}_{\mathbf{k}',\phi\phi}(\eta)$ can be written as
\begin{equation}
	\begin{aligned}
		\langle h^{\lambda,(3)}_{\mathbf{k},\phi \psi_{\phi\phi}}&(\eta) h^{\lambda',(2)}_{\mathbf{k}',\phi\phi}(\eta) \rangle=\int\frac{d^3p}{(2\pi)^{3/2}}\int\frac{d^3q}{(2\pi)^{3/2}}\\
		&\times\int\frac{d^3p'}{(2\pi)^{3/2}}\varepsilon^{\lambda, lm}(\mathbf{k})\varepsilon^{\lambda', rs}(\mathbf{k}')p'_r p'_s p_lp_m \\
		& \times 
		I^{(3)}(u,\bar{u},\bar{v},x)I^{(2)}_{\phi\phi}(u',v',x')\\
		& \times \langle \zeta_{\mathbf{k}-\mathbf{p}} \zeta_{\mathbf{p}-\mathbf{q}} \zeta_{\mathbf{q}}\zeta_{\mathbf{k}'-\mathbf{p}'} \zeta_{\mathbf{p}'} \rangle \ , 
	\end{aligned}\label{eq:P}
\end{equation}
where $|\mathbf{k}'-\mathbf{p}'|=u'|\mathbf{k}'|$, $|\mathbf{p}'|=v'|\mathbf{k}'|$, and $x'=|\mathbf{k}'|\eta$. As shown in Eq.~(\ref{eq:P}), we encounter the five-point correlation function of the primordial curvature perturbation. Here, we consider the lowest order contributions of local-type primordial non-Gaussianity. Specifically, in the five-point correlation function of primordial curvature perturbation, only one perturbation is non-Gaussian, while the other four are Gaussian. In this case, the two-point function $\langle h^{\lambda,(3)}_{\mathbf{k},\phi \psi_{\phi\phi}}(\eta) h^{\lambda',(2)}_{\mathbf{k}',\phi\phi}(\eta)\rangle$ is proportional to $f_{\mathrm{NL}}A_{\zeta}^3$. As shown in diagrams (b) and (c) in Fig.~\ref{fig:FeynDiagLarge}, there are two non-equivalent two-loop diagrams for the case where the non-Gaussianity of the primordial curvature perturbation originates from $h^{\lambda',(2)}_{\mathbf{k}',\phi\phi}(\eta)\sim\zeta_{\mathbf{k}'-\mathbf{p}'}\zeta_{\mathbf{p}'}$ in Eq.~(\ref{eq:P}). Moreover, the diagrams (d)--(f) in  Fig.~\ref{fig:FeynDiagLarge} illustrate that there are three distinct two-loop diagrams corresponding to the non-Gaussianity of the primordial curvature perturbation originating from $h^{\lambda,(3)}_{\mathbf{k},\phi\psi_{\phi\phi}}(\eta)\sim\zeta_{\mathbf{k}-\mathbf{p}}\zeta_{\mathbf{p}-\mathbf{q}}\zeta_{\mathbf{q}}$ in Eq.~(\ref{eq:P}). 

\begin{figure}[htbp]
    \centering
    \includegraphics[width=.3\columnwidth]{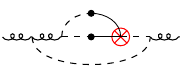}
    \includegraphics[width=.3\columnwidth]{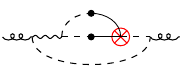}
    \includegraphics[width=.3\columnwidth]{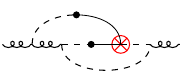} \\
    \includegraphics[width=.3\columnwidth]{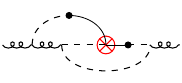}
    \includegraphics[width=.3\columnwidth]{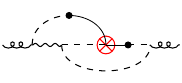}
    \includegraphics[width=.3\columnwidth]{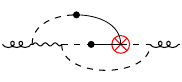} \\
    \includegraphics[width=.3\columnwidth]{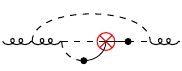}
    \includegraphics[width=.3\columnwidth]{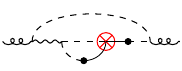}  
    \includegraphics[width=.3\columnwidth]{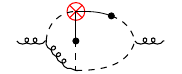} \\
    \includegraphics[width=.3\columnwidth]{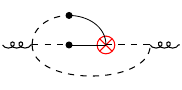} 
    \includegraphics[width=.3\columnwidth]{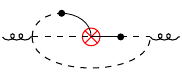} 
    \includegraphics[width=.3\columnwidth]{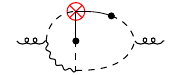}
    
\caption{\label{fig:FeynDiagSmall} The two-loop diagrams correspond to the two-point functions: $\langle h^{\lambda,(3)}_{\mathbf{k},\phi\phi\phi} h^{\lambda',(2)}_{\mathbf{k}',\phi\phi} \rangle$, $\langle h^{\lambda,(3)}_{\mathbf{k},\phi h_{\phi\phi}} h^{\lambda',(2)}_{\mathbf{k}',\phi\phi} \rangle$, and $\langle h^{\lambda,(3)}_{\mathbf{k},\phi V_{\phi\phi}} h^{\lambda',(2)}_{\mathbf{k}',\phi\phi} \rangle$. The dashed lines, wavy lines and spring-like lines represent scalar, vector, and tensor perturbations, respectively. }
\end{figure}
The power spectra $\mathcal{P}^{(m,n)}_k(\eta)$ of the two-point function $\langle h^{\lambda,(m)}_{\mathbf{k}}(\eta) h^{\lambda',(n)}_{\mathbf{k}'}(\eta) \rangle$ are defined as
\begin{equation}\label{eq:Ph} 
	\begin{aligned}
		\langle h^{\lambda,(m)}_{\mathbf{k}}(\eta) h^{\lambda^{\prime},(n)}_{\mathbf{k}'}(\eta)\rangle= \delta^{\lambda \lambda'}\delta\left(\mathbf{k}+\mathbf{k}'\right) \frac{2 \pi^{2}}{k^{3}} \mathcal{P}^{(m,n)}_k(\eta) \ .
	\end{aligned}
\end{equation}
In the spherical coordinate system, we can obtain the explicit expression of the power spectrum that corresponds to Eq.~(\ref{eq:P})
\begin{equation}\label{eq:Pex} 
	\begin{aligned}
		\mathcal{P}^{(3,2)}_{\phi\psi_{\phi\phi},\phi\phi} & = \frac{16f_{\mathrm{NL}}}{405 \pi} \int_0^{\infty} {\rm d} v \int_0^{\infty}{\rm d} \bar{v} \int^{1 + \bar{v} v}_{| 1 -  \bar{v} v |} {\rm d} w \int^{1 + v}_{| 1 - v |}{\rm d} u \\
		&\int_{\bar{u}_-}^{\bar{u}_+} {\rm d} \bar{u} \Big\{ \frac{w}{u^2 \bar{u}^2 v \bar{v}^2 \sqrt{Y (1 - X^2)}}~ \Lambda^{lmrs}(\mathbf{k})p_lp_m   \\
		&  \times I^{(3)}_{\phi\psi_{\phi\phi}} \left( u, v, \bar{u}, \bar{v}, x \right) \sum_{a=1}^{n}\left(p'_r p'_s  I^{(2)}_{\phi\phi} \left( u', v', x \right) \right)_a \\
		&P_{\zeta} (k  u) P_{\zeta} (k \bar{u} v) P_{\zeta}(k \bar{v} v) \Big\} \ ,    
	\end{aligned}
\end{equation}
where $P_{\zeta}$ is the primordial power spectrum which is defined as $\langle \zeta_{\mathbf{k}} \zeta_{\mathbf{k}'}  \rangle=\frac{2\pi^2}{k^3}\delta(\mathbf{k}+\mathbf{k}')\mathcal{P}_{\zeta}(k)$. The subscripts $\phi\psi_{\phi\phi}$ and $\phi\phi$ on the left-hand side of the equation represent the source term of \acp{GW} in the two-point correlation function. The summation of index $a$ in Eq.~(\ref{eq:Pex}) represents the Wick's expansions of the six-point function of $\zeta_{\mathbf{k}}$. The simplification of the integral in Eq.~(\ref{eq:Pex}) is equivalent to that of the third order \acp{SIGW}. The specific form and detailed derivation process of the integral can be found in Refs.~\cite{Chang:2022dhh,Zhou:2021vcw}.

Note that there are four kinds of source terms for the third order \acp{SIGW} in Eq.~(\ref{eq:H0}). The other three kinds of two-point correlation functions can also be investigated in the same way as Eq.~(\ref{eq:P}). Fig.~\ref{fig:FeynDiagSmall} shows the two-loop diagrams corresponding to all remaining two-point correlation functions. 

\begin{figure}[htbp]
    \centering
    \includegraphics[width=.95\columnwidth]{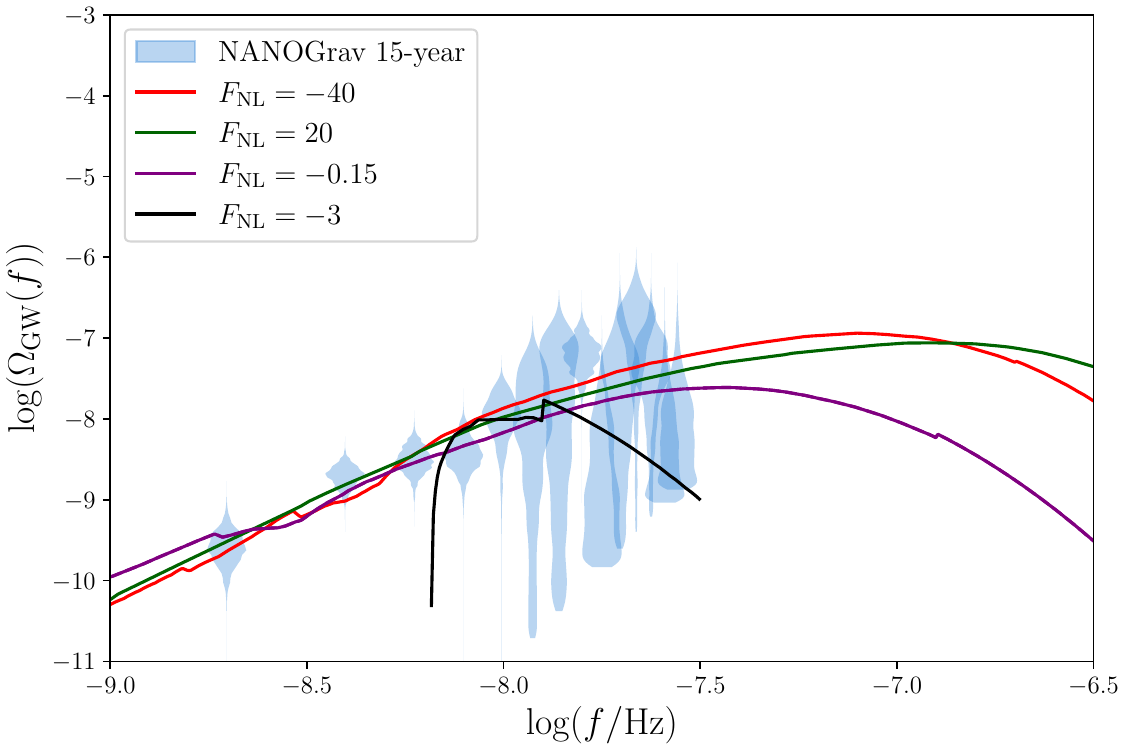}
\caption{\label{fig:violin} The total energy density spectra $\Omega_{\mathrm{GW},0}^{\mathrm{tot}}$ with $\sigma_*=1$ for different $f_{\mathrm{NL}}$. The parameters $\left (\log(A_\zeta),\log(k_*) \right)$ for the red, green, purple, and black solid lines are $(-1.65,-7.3)$, $(-1.55,-7)$, $(-1,-7.5)$, and $(-1.1,-8.5)$, respectively. The energy density spectra derived from the free spectrum of NANOGrav 15-year data set are also shown here.}
\end{figure}

\begin{figure}[htbp]
    \centering
    \includegraphics[width=.95\columnwidth]{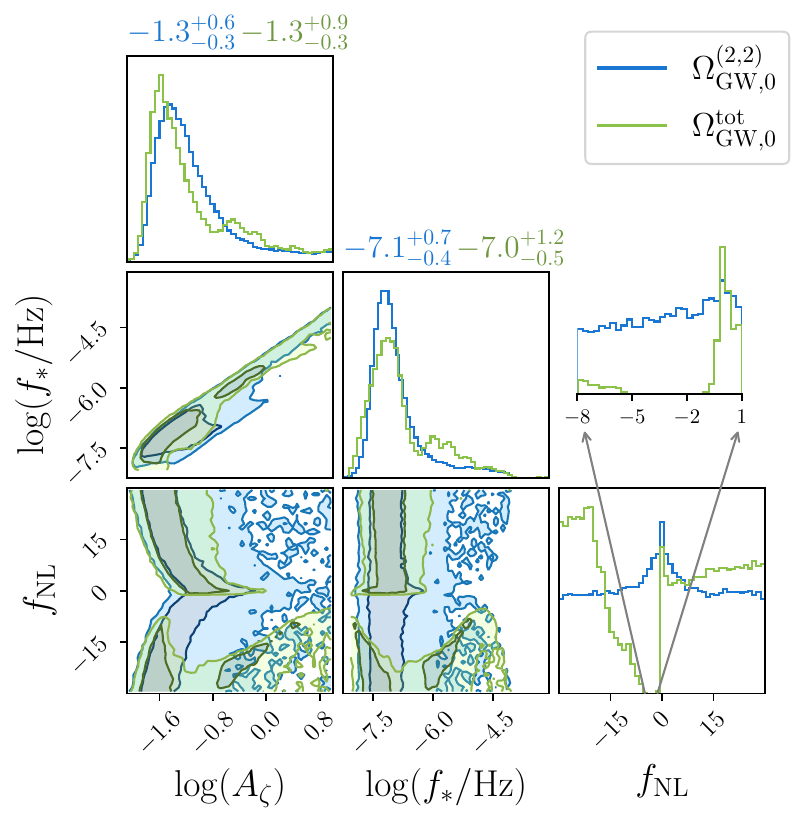}
\caption{\label{fig:corner} One and two-dimensional marginalized posteriors of three independent parameters, $\log (A_{\zeta})$, $\log (f_*)$, and $f_{\mathrm{NL}}$. The priors are $\log (A_{\zeta}) \in [-5, 1]$, $\log (f_*/\mathrm{Hz}) \in [-10,-3]$, and $f_{\mathrm{NL}} \in [-30,30]$. The previous one-loop and two-loop contributions are represented by the blue lines, and the green lines show the result after the addition of the new two-loop contributions.}
\end{figure}

Here, we consider the log-normal primordial power spectrum
\begin{eqnarray}\label{eq:Pr}
	\mathcal{P}_{\zeta}(k)=\frac{A_{\zeta}}{\sqrt{2 \pi \sigma_*^2}} \exp \left(-\frac{\ln^2 \left(k / k_*\right)}{2 \sigma_*^2}\right) \ .
\end{eqnarray}
The total energy density spectrum of \acp{SIGW} is given by 
\begin{equation}\label{eq:Omega}
	\begin{aligned}
		\Omega_{\mathrm{GW}}^{\mathrm{tot}}&=\Omega_{\mathrm{GW}}^{(2,2)}+\Omega_{\mathrm{GW}}^{(2,3)}\\
  &=\frac{1}{6}\left(\frac{k}{a(\eta) H(\eta)}\right)^{2} \left( \frac{1}{4}\mathcal{P}^{(2,2)}_{h}+\frac{1}{6}\mathcal{P}^{(3,2)}_{h} \right) \ .
	\end{aligned}
\end{equation}
Taking into account the thermal history of the universe, we obtain the current total energy density spectrum $\Omega^{\mathrm{tot}}_{\mathrm{GW},0}$ \cite{Wang:2019kaf}
\begin{equation}\label{eq:tot_spectrum}
	\begin{aligned}
	\Omega_{\mathrm{GW},0}^{\mathrm{tot}}=\Omega_{\mathrm{rad}, 0}\left(\frac{g_{*, \rho, \mathrm{e}}}{g_{*, \rho, 0}}\right)\left(\frac{g_{*, s, 0}}{g_{*, s, \mathrm{e}}}\right)^{4 / 3} \bar{\Omega}^{\mathrm{tot}}_{\mathrm{GW}}(\eta, k) \ .
	\end{aligned}
\end{equation}
After considering the contributions of seventeen new two-loop diagrams shown in Fig.~\ref{fig:FeynDiagLarge} and Fig.~\ref{fig:FeynDiagSmall}, we present the current total energy density spectrum in Fig.~\ref{fig:violin}. The power spectrum of \ac{SIGW} originates from three contributions: the Gaussian $\langle h^{\lambda,(2)}_{\mathbf{k}}(\eta) h^{\lambda',(2)}_{\mathbf{k}'}(\eta) \rangle\sim A_{\zeta}^2$ , the non-Gaussian $\langle h^{\lambda,(2)}_{\mathbf{k}}(\eta) h^{\lambda',(2)}_{\mathbf{k}'}(\eta) \rangle\sim (f_{\mathrm{NL}})^2A_{\zeta}^3$, and the non-Gaussian $\langle h^{\lambda,(3)}_{\mathbf{k}}(\eta) h^{\lambda',(2)}_{\mathbf{k}'}(\eta) \rangle\sim f_{\mathrm{NL}}A_{\zeta}^3$. We use Ceffyl \cite{Lamb:2023jls} package embedded in PTArcade \cite{Mitridate:2023oar}  to analyze the data from the first 14 frequency bins of NANOGrav 15-year data set. Since previous work has primarily considered only the contributions from the first two parts, as shown in Fig.~\ref{fig:corner}, the blue curve of $f_{\mathrm{NL}}$ is symmetric about zero. After taking into account the contributions of the two-point correlation function $\langle h^{\lambda,(3)}_{\mathbf{k}}(\eta) h^{\lambda',(2)}_{\mathbf{k}'}(\eta) \rangle$, the green curve of $f_{\mathrm{NL}}$ is no longer symmetric, and the parameter interval $f_{\mathrm{NL}}\in [-5,-1]$ is significantly excluded. When $f_{\mathrm{NL}}\in [-5,-1]$, the contributions of the missing two-loop diagrams will critically suppress the total energy density spectrum (black curve in Fig.~\ref{fig:violin}), which prevents the total energy density spectrum of SIGWs from fitting well with the observational data from NANOGrav 15-year in this parameter interval.

For the \acp{PBH} formation, a positive value of $f_{\mathrm{NL}}$ will augment the abundance of \acp{PBH} for a given power spectrum of curvature perturbation. Conversely, a negative value of $f_{\mathrm{NL}}$ will reduce the abundance of \acp{PBH} \cite{Young:2013oia}. To avoid the overproduction of \acp{PBH}, it is generally necessary to consider the case where $f_{\mathrm{NL}}<0$, since the abundance of PBHs cannot exceed that of dark matter \cite{Carr:2020xqk}. When considering the constraints from \ac{PTA} observations and the abundance of primordial black holes together, the possible range for $f_{\mathrm{NL}}$ is $ f_{\mathrm{NL}}<-5$ or $-1\le f_{\mathrm{NL}}< 0$. Since the constraints of primordial non-Gaussianity on small scales ($\lesssim$1 Mpc) are significantly weaker than those on large scales \cite{Planck:2018vyg,Abdalla:2022yfr,Bringmann:2011ut}, $|f_{\mathrm{NL}}|$ could be substantially greater on small scales. It seems that the parameter interval $ f_{\mathrm{NL}}<-5$ cannot be ruled out yet. However, to ensure the convergence of cosmological perturbations, $f_{\mathrm{NL}}$ and $A_{\zeta}$ must satisfy $(f_{\mathrm{NL}})^2A_{\zeta}<1$ and $A_{\zeta}<1$ in the non-Gaussian scenario. Fig.~\ref{fig:corner} shows that $\log (A_{\zeta})=(-1.3)^{+0.9}_{-0.3}$ with $1-\sigma$ credible region, implying that the absolute value of $f_{\mathrm{NL}}$ is less than $\sqrt{1/A_{\zeta}}\approx 4.47$. Consequently, even though the results with $f_{\mathrm{NL}}=-40$ in Fig.~\ref{fig:violin} can fit the NANOGrav 15-year data well, the region of the parameter space $f_{\mathrm{NL}}<-5$ might be ruled out by the convergence of perturbation expansion. Therefore, after taking into account the upper limit of the abundance of \acp{PBH} and the convergence of the cosmological perturbation expansion, we find that the only possible range for $f_{\mathrm{NL}}$ might be $-1\le f_{\mathrm{NL}}< 0$ (purple line in Fig.~\ref{fig:violin}).

\emph{\textbf{Conclusion.}}---In this Letter, we systematically studied the seventeen missing two-loop diagrams proportional to $f_{\mathrm{NL}}A_{\zeta}^3$ that correspond to the two-point correlation function $\langle h^{\lambda,(3)}_{\mathbf{k}} h^{\lambda',(2)}_{\mathbf{k}'} \rangle$ in the case of local-type primordial non-Gaussianity. The missing two-loop contributions significantly impact the total energy density spectrum of the \acp{SIGW}.

When combined with observations from the NANOGrav 15-year data, we found that if \acp{SIGW} dominate the \acp{SGWB} observed in \ac{PTA} experiments, the parameter range for $ f_{\mathrm{NL}} \in [-5, -1]$ will be entirely excluded. Furthermore, after taking into account the upper limit of the abundance of \acp{PBH} and the convergence of the cosmological perturbation expansion, we conclude that the parameter spaces of $f_{\mathrm{NL}}\geq 0$ and $f_{\mathrm{NL}}<\sqrt{1/A_{\zeta}}\approx-4.47$ are ruled out, the only possible range for $f_{\mathrm{NL}}$ might be $-1\le f_{\mathrm{NL}}< 0$.

\vspace{0.3cm}
\begin{acknowledgements} 
The authors want to thank Quan-feng Wu for useful discussions and valuable suggestions. This work has been funded by the National Nature Science Foundation of China under grant No. 12075249 and 11690022, and the Key Research Program of the Chinese Academy of Sciences under Grant No. XDPB15. We acknowledge the \texttt{xPand} package \cite{Pitrou:2013hga}. 
\end{acknowledgements}

\bibliography{Two-loop}

\end{document}